\documentclass[twocolumn,showpacs,preprintnumbers,amsmath,amssymb,nofootinbib]{revtex4}
\usepackage{latexsym}%
\usepackage{graphicx}
\usepackage{dcolumn}
\usepackage{bm}

\begin{document}

\voffset= 1.0 truecm 

%
%
%
%
%
%
%
%
%
%
%
%
%
%
%
%
%
%
%

%
\newcommand{\bea}{\begin{eqnarray}}
\newcommand{\eea}{\end{eqnarray}}
\newcommand{\be}{\begin{equation}}
\newcommand{\ee}{\end{equation}}
%
\newcommand{\xbf}[1]{\mbox{\boldmath $ #1 $}}
%
\newcounter{saveeqn}
\newcommand{\alpheqn}{\setcounter{saveeqn}{\value{equation}}%
\setcounter{equation}{0}%
\renewcommand{\theequation}{\mbox{\arabic{saveeqn}\alph{equation}}}}
\newcommand{\reseteqn}{\setcounter{equation}{\value{saveeqn}}%
\renewcommand{\theequation}{\arabic{equation}}}
%
%
\def\shiftleft#1{#1\llap{#1\hskip 0.04em}}
\def\shiftdown#1{#1\llap{\lower.04ex\hbox{#1}}}
\def\thick#1{\shiftdown{\shiftleft{#1}}}
\def\b#1{\thick{\hbox{$#1$}}}

\title{Quadrupole moments of baryons}
\thanks{Published in Phys. Rev. D {\bf 65}, 073017 (2002).}
\author{A. J. Buchmann$^{1}$ and E. M. Henley$^{2}$}
\affiliation{$^1$ Institut f\"ur Theoretische Physik, Universit\"at T\"ubingen\\
Auf der Morgenstelle 14, D-72076 T\"ubingen, Germany \\
$^2$ Department of Physics and Institute for Nuclear Theory, Box 351560, \\ 
University of Washington, Seattle, WA 98195, U.S.A.\\ }

\begin{abstract}
Quadrupole moments of decuplet baryons and the octet-decuplet
transition quadrupole moments are calculated using Morpurgo's
general QCD parameterization method. Certain relations 
among the decuplet and the octet to decuplet transition
quadrupole moments are derived. 
These can be used to predict 
the $\Delta$ quadrupole moments which are difficult to measure.
\end{abstract}


\maketitle


\section{\bf Introduction}
We use Morpurgo's QCD parameterization~\cite{Mor89}  
to calculate all baryon octet-decuplet transition 
and decuplet quadrupole moments and to derive certain relations 
between them. The chosen method 
makes it clear from the outset that the results obtained do not depend on the 
particular quark-quark interaction model. We compare our relations
with those obtained in a group theoretical analysis. 
Finally, we
provide numerical estimates that can be compared with experiment.

Unfortunately, quadrupole moments of the decuplet baryons are still 
unknown. They are important for providing evidence for baryon 
non-sphericity. We believe that the electric quadrupole moment of the 
$\Omega^-$, as well as the octet-decuplet transition quadrupole moments,
are amenable to measurement and we discuss some possible techniques 
in the summary. 

\section{Morpurgo's general parameterization method for quadrupole moments}
The general parameterization (GP) method, developed by Morpurgo 
is based on the symmetries and the quark-gluon dynamics of the 
underlying field theory of quantum chromodynamics (QCD). Although 
noncovariant in appearance, all invariants that are allowed by Lorentz 
invariance are included in the operator basis (see below).

The basic idea is to {\it formally} define, for the observable at 
hand, a QCD operator $\Omega$ and QCD eigenstates $\vert B \rangle$  
expressed explicitly
in terms of quarks and gluons. The corresponding matrix element
can, with the help of the unitary operator $V$, be reduced to an
evaluation in the basis of auxiliary three-quark states 
$\vert\Phi_B \rangle $
\begin{equation}
\label{map}
\left \langle B \vert \Omega \vert B \right \rangle =
\left \langle \Phi_B \vert
V^{\dagger}\Omega V \vert \Phi_B \right \rangle =
\left \langle W_B \vert
{ {\cal O}} \vert W_B \right \rangle \, .
\end{equation}
The auxiliary states $\vert \Phi_B \rangle$ 
are pure three-quark states with orbital angular momentum
$L=0$.  The spin-flavor wave functions~\cite{Lic78} 
contained in $\vert \Phi_B \rangle$ are denoted by $\vert W_B\rangle $. 
The operator $V$ dresses the pure three-quark 
states with $q\bar q$ components and gluons and 
thereby generates
the exact QCD eigenstates $\vert B \rangle $. Furthermore, 
it is implied that $V$
contains a Foldy-Wouthuysen transformation allowing the 
auxiliary states to be written in terms of Pauli spinors.

One then writes the most general expression for the operator 
${ {\cal O}}$, in the present case for the electric 
quadrupole operator ${ {\cal Q}}$, that is 
compatible with the space-time and inner QCD symmetries.
The orbital and color space 
matrix elements~\footnote{Note that on the right hand side of the last equality 
in Eq.(\ref{map}) the integration over spatial and color degrees of 
freedom has been performed.} are absorbed
into {\it a priori} unknown parameters, called B and C, multiplying 
the spin-flavor invariants
appearing in the expansion of ${ {\cal O}}$. The method has
been used to calculate various properties of 
baryons and mesons~\cite{Mor89,Mor99}. 

The electric quadrupole operator is composed of a two- and three-body term
in spin-flavor space
\bea
\label{para1}
{ {\cal Q}} &= & B\sum_{i \ne j}^3 e_i \left ( 3 \sigma_{i \, z} \sigma_{ j \, z}
-\b{\sigma}_i \cdot \b{\sigma}_j \right ) \nonumber \\
& + & C \sum_{i \ne j \ne k }^3 e_k  
\left ( 3 \sigma_{i \, z} \sigma_{ j \, z} - 
\b{\sigma}_i \cdot \b{\sigma}_j \right ), 
\eea
where 
$e_i=(1 + 3 \tau_{i \, z})/6$ is the charge of the i-th quark.
More general operators containing second and third 
powers of the quark charge are conceivable~\cite{Leb00} but are 
not considered here. Their contribution is suppressed by factors 
of $e^2/4\pi=1/137$. The $z$-component
of the Pauli spin (isospin) matrix $\b{\sigma}_i$ ($\b{\tau}_i$) 
 is  denoted by $\sigma_{i \, z}$ ($\tau_{i \, z}$).

Decuplet baryon quadrupole moments $Q_{B^*}$ and octet-decuplet transition 
quadrupole moments $Q_{B \to B^*}$ are obtained by calculating the
matrix elements of the quadrupole operator 
in Eq.(\ref{para1}) between the 
three-quark spin-flavor wave functions $\vert W_B \rangle $
\begin{eqnarray}
\label{matrixelements} 
Q_{B^*} & = &\left \langle W_{B^*} \vert { {\cal Q}} 
\vert W_{B^*} \right \rangle , \nonumber \\
Q_{B \to B^*} 
& = & \left \langle W_{B^*} \vert {{\cal Q}} \vert W_B \right \rangle,  
\end{eqnarray}  
where $B$  denotes a spin 1/2 octet baryon and $B^*$ a member of the
spin 3/2 baryon decuplet.

\subsection{Missing one-quark quadrupole operator}
In applications of the GP, a hierachy in the importance 
of one-, two-, and three-body operators 
is often found. One-body operators usually give a larger contribution 
to the matrix element than two-body 
operators, and two-body operators
are usually more important than three-body operators \cite{Mor89}. 
This hierarchy results
from the additional gluon exchanges needed to generate two-quark and 
three-quark operators. 
The quark-gluon coupling $\alpha_s=g^2/4\pi$ is such that
diagrams involving higher powers of $g$ are suppressed.
In the GP method this is regarded as an empirical fact; 
in QCD with a large number ($N_c$) of colors this is because 
$g$ is inversely proportional to $N_c$, i.e., 
$g \propto  1/\sqrt{N_c}$, and diagrams involving
higher powers of $g$ are suppressed.
 
However, for quadrupole moments,
one-body operators containing  rank 2 spherical harmonics in orbital space, 
e.g., $e_i \, Y^2({\bf {\hat x}})$,
do not contribute, because the GP method employs only
$L=0$ wave functions. 
In Morpurgo's formulation, any possible $D$-state admixture in the QCD-states 
$\vert B \rangle$ is moved from the wave function to the effective two- 
and three-quark operators ${\cal Q}$ acting in spin-flavor space.
Even if $D$-waves were included in the Hilbert space, 
the orbital one-body contribution would be small due to 
the small $D$-state probability 
in the nucleon and $\Delta$ wave functions~\cite{Ger82}. 
In any case, for quadrupole moments, 
we are left with two- and three-quark operators.
\begin{table*}[htb]
\begin{center}
\begin{tabular}{ l | r |  r | r } 
\hline
\hline
& $Q(r=1)$  & $Q$(quadratic)   & $Q$(cubic)  \\
\hline
$\Delta^{-}$     & $ -4B -4C$	  & $ -4B -4C$        & $-4B -4C$           \\
$\Delta^{0}$     &     0     	  &     0             & 0                   \\
$\Delta^{+}$     & $4B + 4C $ 	  & $4B + 4C $        & $4B +4C$            \\
$\Delta^{++}$    & $8B + 8C$  	  & $8B + 8C$         & $8B +8C$            \\
\hline
 & &   \\
$\Sigma^{\ast -}$ & $-4B-4C$  &          $-(4B + 4C ) (1 +2r)/3  $ 
& $-(4B+4C) (1+r+r^2)/3 $      \\
$\Sigma^{\ast 0}$ & 0    
& $2 (B-2C) (1 -r)/3$  & $ [2B (1+r-2r^2) - 2C(2-r-r^2)]/3$   \\    
$\Sigma^{\ast +}$ & $4B+4C$   
& $[4B (2 +r) - 4 C (1-4r)]/3 $  & $ [4B(2 + 2r -r^2) -4C(1-2r-2r^2)]/3 $  \\ 
\hline
  & & \\
$\Xi^{\ast -}$ & $-4B-4C$      
& $-(4B + 4C)(2r +r^2)/3 $ & $-(4B+4C)(r + r^2 +r^3)/3$     \\
$\Xi^{\ast 0}$ &  0 
& $4 (B - 2C) (r -r^2)/3 $  & $[4B(2r-r^2-r^3) -4C(r+r^2-2r^3)]/3$ \\ \hline  
  & & \\
$\Omega^-$ & $-4B-4C$	
& $-(4B + 4C)  r^2  $ & $-(4B + 4C)r^3 $          \\  
\hline
\hline
\end{tabular} 
\caption[C2 moments]{Two-quark ($B$) and three-quark ($C$) contributions to 
quadrupole moments of decuplet baryons 
in the SU(3) symmetry limit ($r=1$)  
and with broken flavor symmetry.
SU(3)-flavor symmetry breaking is characterized by the ratio of 
u-quark and s-quark masses $r=m_u/m_s$.
Two types (quadratic and cubic) of flavor symmetry
breaking are considered. } 
\label{quadmo}
\end{center}
\end{table*}

\subsection{Two- and three-quark quadrupole operators}
The two- and three-body operators in Eq.(\ref{para1}) act 
in spin-flavor space. Although they
formally operate  on valence quark states, they are mainly 
a reflection of the $q \bar q$ and gluon 
degrees of freedom that have been
eliminated from the Hilbert space, and which reappear as a quadrupole
tensor in spin space~\cite{Hen01,Buc97}. 
As spin tensors of rank 2, they can induce 
spin $1/2 \to 3/2$ and $3/2 \to 3/2$ transitions.  

Evaluating Eq.(\ref{para1})
between, e.g., $N$ and $\Delta$ spin-isospin wave functions leads to
the following results for the $\Delta$ and the $N \to  \Delta$ 
quadrupole moments 
\bea
\label{twothree}
Q_{\Delta^+} & = &\langle W_{\Delta^+} 
\vert { {\cal Q}}_{[2]} + { {\cal Q}}_{[3]} 
\vert  W_{\Delta^+} \rangle  =  4 B + 4 C,  \nonumber \\
Q_{p \to \Delta^+} & = & \langle W_{\Delta^+} 
\vert { {\cal Q}}_{[2]} + { {\cal Q}}_{[3]} 
\vert  W_{p} \rangle  =  2 \sqrt{2} (B \!- \! 2C).  
\eea
Similarly, the electric quadrupole moments for the other decuplet baryons
and the octet-decuplet transition moments 
are calculated and listed in Tables \ref{quadmo} and \ref{transquad}.
In this way Morpurgo's method yields an efficient parameterization
of baryon quadrupole moments in terms of few unknown parameters.
\begin{table*}[htb]
\begin{center}
\begin{tabular}{ l |  r | r | r  } 
\hline
\hline
&  $Q(r=1)$ &   $Q$(quadratic) & $Q$(cubic)  
\\ \hline 
$p\to \Delta^+$  & $2\sqrt{2} (B-2C)$  & $2\sqrt{2} \,  (B-2C)$  
& $2\sqrt{2} \,  (B-2C)$      \\
$n\to \Delta^0$ & $2\sqrt{2} (B-2C)$ &  $2\sqrt{2} \, (B-2C)$ & $2\sqrt{2} \,  
(B-2C)$       \\
\hline
 & &   \\
$\Sigma^- \to \Sigma^{\ast -}$  & 0 & 
$-2\sqrt{2} \, (2B+2C)\, (1-r)/3$ & 
$-\sqrt{2} \, (2B+2C)\, (2-r-r^2)/3$    \\
$\Sigma^0 \to \Sigma^{\ast 0}$ &  $\sqrt{2}(B-2C)$ &
$\sqrt{2}\,  (B-2C) \,  (2+r)/3$ &  
$\sqrt{2} [2B (2-r+2r^2) - 2C (4 + r +r^2)]/6 $ \\
$\Lambda^0 \to \Sigma^{\ast 0}$ & $\sqrt{6} (B-2C)$ & 
$\sqrt{6} \,(B-2C)\, r $ & 
$\sqrt{6}[2B r - 2C  (r + r^2)]/2 $  \\
$\Sigma^+ \to \Sigma^{\ast +}$ & $2\sqrt{2} (B-2C)$ & 
 $2\sqrt{2}\, \lbrack B \,(4-r) - 2C \,(1+2r) \rbrack /3 $   & 
 $2\sqrt{2}\, \lbrack B \,(4-2r+r^2) - 2C \,(1+r +r^2) \rbrack /3 $   
\\
\hline
 & &   \\
$\Xi^- \to \Xi^{\ast -}$ & 0 & 
   $-2\sqrt{2} \,(2B+2C)\, (r-r^2)/3$     & 
   $-\sqrt{2} \,(2B+2C)\, (r+r^2-2r^3)/3$     
\\
$\Xi^0 \to \Xi^{\ast 0}$ & $2\sqrt{2} (B-2C)$ & 
    $2\sqrt{2}\, (B-2C) \,(r+2r^2)/3$  &
   $\sqrt{2}[2B (2r -r^2 + 2r^3) - 2C (r + r^2 + 4r^3)]/3$    
\\
\hline
\hline
\end{tabular} 
\caption[C2 Transition moments]{Two-quark ($B$) and three-quark ($C$) 
contributions 
to the octet-decuplet transition quadrupole moments
in the SU(3) symmetry limit ($r=1$)
and with broken flavor symmetry.
SU(3)-flavor symmetry breaking is characterized by the ratio of 
u-quark and s-quark masses $r=m_u/m_s$.
Two types (quadratic and cubic) of flavor symmetry
breaking are considered. } 
\label{transquad}
\end{center}
\end{table*}
\subsection{Determination of the GP constants}
In order to  determine the two constants $B$ and $C$ we need two 
experimental inputs.  From recent measurements of the ratio of 
electric quadrupole over magnetic dipole amplitudes 
in electromagnetic pionproduction 
($E2/M1$ and $C2/M1$ ratios)~\cite{Bla97,Bec97}, one can 
extract the $N \to \Delta $ transition quadrupole moment 
$Q_{p \to \Delta^+}$. For a first determination of $Q_{\Delta^+}$ from 
photo-pionproduction data in the $\Delta$ resonance region 
see Ref.~\cite{Bla97}. Because the decuplet quadrupole moments  
or other octet-decuplet transition quadrupole moments are not yet very 
well known, we cannot fix the smaller constant $C$ with sufficient 
accuracy at this stage. Therefore, we assume $C \approx 0$ for the numerical
evaluation.  Our assumption 
that three-body (C) terms in the charge operator 
are smaller than two-body (B) terms is supported  by  
work using the GP~\cite{Mor99} and the $1/N_c$ expansion~\cite{Leb00} methods. 
In both methods, $\vert C/B  \vert$ is estimated to be at most 0.3.

We take the following approach in determining the constant $B$.
In a quark model with exchange currents, 
it was found that the $N \to \Delta$
and $\Delta$ quadrupole moments receive the largest contribution
from two-body $q \bar q$ terms in the charge operator.
The following relations between the neutron charge radius $r_n^2$, 
the $\Delta^+$, and the $p \to \Delta^+$ quadrupole moments were 
obtained~\cite{Buc97} 
\begin{eqnarray}
\label{relations}
\sqrt{2} \, Q_{p \to \Delta^+} &  = &  
Q_{\Delta^+} = r_n^2.
\end{eqnarray}
The reasons for the existence of these quark model relations are:
(i) one-quark and three-quark operators are suppressed 
for $r_n^2$, $Q_{\Delta^+}$, 
and $Q_{p \to \Delta^+}$, 
(ii) these observables are dominated by the same two-body charge operator
$\rho_{[2]}$, 
(iii) there is a definite relation between the 
monopole (C0) term in $\rho_{[2]}$, which is responsible 
for the nonvanishing 
neutron charge radius, and the quadrupole (C2) term in $\rho_{[2]}$ 
that produces a nonzero quadrupole moment of the $\Delta^+$.
In other words, all three observables are  
dominated by the cloud of $q \bar q$ pairs 
--effectively described by the two-quark exchange currents--
and as a consequence are simply related.

Similar relations between $r_n^2$, $Q_{\Delta^+}$, and  
$Q_{p \to \Delta^+}$ were
obtained in the pion cloud model~\cite{Hen01} where 
the $q \bar q$ degrees of freedom enter in terms 
of an explicit pion contribution to the baryon wave functions,
and in previous quark model calculations~\cite{Ric84}.

From Eq.(\ref{twothree}) it is clear that 
the quark model relation between $Q_{\Delta^+}$ and $Q_{p \to \Delta^+}$
can only be exact if $C=0$. This is not the case.
Nevertheless, it 
is well satisfied by the experimental neutron
charge radius and the $p \to \Delta^+$ quadrupole moment extracted from 
the electromagnetic pionproduction data~\cite{Bla97,Bec97,Gra01}. This
provides some experimental evidence for the smallness of the constant $C$.
By comparing Eq.(\ref{twothree}) for the case $C=0$  
with  Eq.(\ref{relations}), we obtain 
$$B=r_n^2/4.$$

\section{Results for Spectroscopic Quadrupole Moments} 
In this section we present our results for baryon quadrupole moments
and interpret them in terms of the higher SU(6) spin-flavor symmetry group and
its SU(3) flavor and SU(2) isospin subgroups. 
The spin-flavor symmetry combines the spin 1/2
flavor octet baryons and the spin 3/2 flavor decuplet baryons 
into the symmetric ${\bf 56}$  dimensional representation of the 
SU(6) spin-flavor group.  If the spin-flavor symmetry was exact, 
octet and decuplet masses would be  equal, 
the charge radii of neutral baryons would be zero, and all 
baryon quadrupole moments would vanish. 
In particular, $M_{\Delta^+}=M_p$, $r_{\Delta^0}^2=r_{n}^2=0$, 
and $Q_{\Delta^+}=Q_{p\to \Delta^+}=0$~\cite{comment1}.

\subsection{SU(6) spin-flavor symmetry breaking}
SU(6) symmetry is only approximately 
realized in nature. It is broken by spin-dependent terms in the strong 
interaction Hamiltonian. Their presence explains why decuplet baryons 
are heavier than their octet member counterparts with the same strangeness. 
Likewise, it is broken by the spin-dependent electric quadrupole 
operators in Eq.(\ref{para1}). These have different matrix elements for 
spin 1/2 octet and spin 3/2 decuplet baryons,
and give rise to nonzero quadrupole moments for decuplet baryons. 
 
In the first column ($r=1)$ of Table~\ref{quadmo} 
we show our results for the decuplet quadrupole moments 
in terms of the GP constants $B$ and $C$ describing the contribution
of two- and three-quark operators,
assuming that SU(3)-flavor symmetry is exact. 
Table~\ref{transquad} lists the corresponding expressions
for the octet-decuplet quadrupole transition moments. 
We observe that the decuplet quadrupole moments are proportional to their 
charge, and that the octet-decuplet transition moments between 
the negatively charged baryons are zero. Both results follow from the
assumed flavor symmetry of the strong interaction. 

\subsection{SU(3) flavor symmetry breaking}
In order to get an idea of the degree of SU(3) flavor symmetry breaking
induced by the electromagnetic transition operator,
we replace the spin-spin terms in Eq.(\ref{para1}) by 
expressions with a ``quadratic'' quark mass 
dependence
$$  
\sigma_{i} \sigma_{j} \rightarrow \sigma_{i} 
\sigma_{j}m_u^2/(m_i m_j)
$$
as obtained from a one-gluon exchange interaction between
the quarks. Flavor symmetry breaking is then characterized by the ratio
$r=m_u/m_s$ of $u$ and $s$ quark masses, which is a known number.
We use the same mass for $u$ and $d$ quarks 
to preserve the SU(2) isospin symmetry of the strong interaction, 
that is known to hold to a very good accuracy.

For comparison, we also use a flavor symmetry breaking of ``cubic'' 
quark mass dependence 
$$
\sigma_{i} \sigma_{j} \rightarrow \sigma_{i} \sigma_{j}m_u^3/(m_i^2 m_j),
$$
that follows from the two-body gluon 
exchange charge density~\cite{comment2}.
This leads to expressions for $Q_{B^*}$ and $Q_{B\to B^*}$ 
containing terms up to third order in $r$. No additional 
parameters are introduced in this way.

We emphasize that this treatment is not exact. The GP method of including 
SU(3) symmetry breaking is to introduce additional operators and
parameters, which guarantees that flavor symmetry breaking is incorporated 
to all orders~\cite{Mor99a}. 
There are then so many undetermined constants that the theory can no 
longer make predictions. We expect that our approximate treatment
includes the most important physical effect.

In the second and third columns of Tables~\ref{quadmo} and \ref{transquad} 
we present the analytic expressions for the decuplet and the 
octet-decuplet transition quadrupole moments with quadratic or cubic 
type of flavor symmetry breaking taken into account.

\subsection{Relations among quadrupole moments}
Even though the SU(6) and SU(3) symmetries are broken, 
there exist --as a consequence of the underlying unitary symmetries---
certain relations among the quadrupole moments.
A relation is the stronger the weaker the assumptions required for its
derivation. We are therefore interested in those relations that hold even
when SU(3) symmetry breaking is included in the charge quadrupole operator. 
These are the ones,  which are most likely satisfied in nature. 
The 18 quadrupole moments (10 diagonal 
decuplet and 8 decuplet-octet transition quadrupole moments) are expressed 
in terms of only two constants $B$ and $C$. Therefore, there
must be 16 relations between them. Given the analytical expressions in 
Tables \ref{quadmo} and \ref{transquad}, it is straightforward to verify 
that the following relations hold 
\setcounter{equation}{6}
\alpheqn
\begin{eqnarray}
\label{rel6a}
0 & = & Q_{\Delta^{-}} + Q_{\Delta^+}, \\
\label{rel6b}
0 & = & Q_{\Delta^{0}}, \\
\label{rel6c}
0 & = & 2\, Q_{\Delta^{-}} + Q_{\Delta^{++}}, \\
\label{rel6d}
0 & = & Q_{\Sigma^{* -}} - 2\, Q_{\Sigma^{* 0}} + Q_{\Sigma^{* +}} , \\
\label{rel6e}
0 & = & 3 ( Q_{\Xi^{* -}} - Q_{\Sigma^{* -}} ) -
( Q_{\Omega^-}- Q_{\Delta^-}), \\
\label{rel6f}
0 & = & Q_{p \to \Delta^{+}} - \, Q_{n \to \Delta^{0} }, \\ 
\label{rel6g}
0 & = & Q_{\Sigma^{-} \to \Sigma^{* -}} - 2 \, Q_{\Sigma^{0} \to \Sigma^{* 0}} 
 + Q_{\Sigma^{ +} \to \Sigma^{* +}}, \\
\label{rel6h}
0 & = &  Q_{\Delta^-} -  Q_{\Sigma^{* -}} 
-\sqrt{2} \, Q_{\Sigma^{-} \to \Sigma^{* -}}, \\ 
\label{rel6i}
0 & = &  Q_{\Delta^+} - Q_{\Sigma^{* +}} + \sqrt{2} \,  
Q_{p \to \Delta^{+}} 
- \sqrt{2} \, Q_{\Sigma^{+} \to \Sigma^{* +}}, \\ 
\label{rel6j}
0 & = & Q_{\Sigma^{* 0}} -\frac{1}{\sqrt{2}} \, Q_{\Sigma^{0} \to \Sigma^{* 0}}
+\frac{1}{\sqrt{6}} Q_{\Lambda^{0} \to \Sigma^{* 0}}, \\
\label{rel6k}
0 & = & Q_{\Sigma^{* -}}\! \!- \!Q_{\Xi^{* -}}\!\! - \!  
\frac{1}{\sqrt{2}} \,   Q_{\Xi^{ -} \to \Xi^{* -}} \!\!- \!
\frac{1}{\sqrt{2}} \,   Q_{\Sigma^{- } \Sigma^{* -}},\\
\label{rel6l}
0 & = &  Q_{\Xi^{* 0}} + \frac{1}{\sqrt{2}} \,  
Q_{\Xi^{0} \to \Xi^{* 0}} - \sqrt{\frac{2}{3}} \, Q_{\Lambda^{0} \to 
\Sigma^{* 0}}.
\end{eqnarray}
These twelve combinations of quadrupole moments do not depend
on the flavor symmetry breaking parameter $r$. 
In fact, Eqs.(\ref{rel6a}-\ref{rel6d}) are already
a consequence of the assumed SU(2) isospin symmetry of the strong interaction,
and hold irrespective of the order of SU(3) symmetry breaking. 
Eq.(\ref{rel6e}) is the quadrupole moment counterpart of  
the ``equal spacing rule'' for decuplet masses. The latter was obtained 
from considering SU(3) invariance of the strong interaction with a second 
order symmetry breaking perturbation~\cite{Oku63}. The remaining relations 
connect states in the octet and decuplet, 
and the assumption of SU(6) symmetric spin-flavor wave 
functions $\vert W_B \rangle$ in the auxiliary states 
is needed to derive them. 

There are also four $r$-dependent 
relations\footnote{The parameter combinations 
$Br$, $Cr$, $Br^2$, and $Cr^2$,
when expressed in terms of quadrupole moments lead to four $r$-dependent
relations among the quadrupole moments. 
We thank R. Lebed for pointing out the proper number
of $r$-dependent relations.}, which can be chosen as 
\setcounter{equation}{7}
\alpheqn
\begin{eqnarray}
\label{rel7a}
0 & = & \frac{1}{3} (2r+1)\, Q_{\Delta^+} + Q_{\Sigma^{* -}},  \\
\label{rel7b}
0 & = & \frac{1}{6}\sqrt{2}(r-1)\, Q_{p \to \Delta^+} +Q_{\Sigma^{* 0}}, \\
\label{rel7c}
0 & = & \sqrt{2}  r^2 \, Q_{p \to \Delta^{+}} 
- \sqrt{2} \,  Q_{\Xi^{0} \to  \Xi^{* 0}} 
+Q_{\Xi^{* 0}}, \\ 
\label{rel7d}
0 & = & r^2 \, Q_{\Delta^-} - Q_{\Omega^-}.
\end{eqnarray}

With the ``cubic'' SU(3) symmetry breaking, 
we obtain the same relations as in Eq.(\ref{rel6a}-\ref{rel6l}) 
with the exception of Eqs.(\ref{rel6j}) and (\ref{rel6l}) 
involving the neutral baryons~\cite{comment3}. 
These no longer hold independently 
but their sum is again a valid relation. 
There are now five $r$-dependent 
relations which can be chosen as
\setcounter{equation}{8}
\alpheqn
\begin{eqnarray}
\label{rel8a}
0 & = & \frac{1}{3}(1+r+r^2) Q_{\Delta^+} + Q_{\Sigma^{* -}}, \\ 
\label{rel8b}
0 & = & (r-r^2) \, Q_{\Delta^+} - \sqrt{2} (2+r^2) Q_{p \to \Delta^{+}} 
\nonumber \\
&+& 6 \sqrt{2} Q_{\Sigma^{0} \to \Sigma^{* 0}}, \\
\label{rel8c}
0 & = &  r \, Q_{\Sigma^{* -}} - Q_{\Xi^{* -}}, \\
\label{rel8d}
0 & = & (r-r^2) Q_{\Delta^+} + \sqrt{2} (r + 2r^3) \, Q_{p \to \Delta^{+}}  \nonumber  \\
&-& 3 \sqrt{2} \,  Q_{\Xi^{0} \to \Xi^{* 0}},\\
\label{rel8e}
0 & = & r^3 \, Q_{\Delta^-} -Q_{\Omega^-}.
\end{eqnarray}
\reseteqn 
Other combinations of the expressions in 
Tables~\ref{quadmo} and \ref{transquad}  can be written down if desirable.

\subsection{Comparison with Lebed's quadrupole moment relations}
It is interesting to compare our results with a pure
SU(6) symmetry analysis of baryon quadrupole moments~\cite{Leb95}. 
After decomposing the 
product ${\bf \overline{56}} \otimes {\bf 56}={\bf 3136}=
{\bf 1} \oplus {\bf 35} \oplus {\bf 405} \oplus {\bf 2695} $
into its irreducible representations, the most general quadrupole 
operator ${\cal Q}$ is expressed in terms of operators 
transforming according to the $ {\bf 405}$ and $ {\bf 2695} $ 
dimensional representations of SU(6) spin-flavor symmetry:
\be
\label{group}
{\cal Q}_{}={\cal Q}_{{\bf 405}} +{\cal Q}_{{\bf 2695}}.
\ee
The ${\bf 1}$ and ${\bf 35}$ 
dimensional representations, which contain only zero-body (constants) and 
one-body operators do not contribute for an angular momentum $J=2$ operator 
such as ${\cal Q}$. 
Lebed's twelve quadrupole moment relations~\cite{Leb95} were derived by 
neglecting the $Q_{\bf 2695}$ terms.
The latter give numerically small 
matrix elements because they require a product of three SU(6) symmetry 
breaking operators, and are therefore suppressed. Omitting 
the $Q_{\bf 2695}$ operators amounts to neglecting the 
three-quark terms $Q_{[3]}$ in the present approach. 

Our results in Tables~\ref{quadmo} and \ref{transquad} with only two-quark
(B) terms retained satisfy all twelve Lebed relations independent 
of whether and how SU(3)-flavor symmetry is broken~\cite{comment4}. 
This shows that
our analytic expressions for the electric quadrupole moments 
are compatible with a rigorous group theoretical approach.

When we include three-body ($C$) operators but no SU(3) symmetry breaking, 
ten of his relations are still satisfied, 
whereas the (8,0), (8,1)\footnote{The first 
number stands for the dimension of the irreducible SU(3)-flavor 
representation, and the second for the isospin $I$ and  
the dimension $2I+1$ of the particular 
SU(2)-isospin representation involved.} relations are violated. 
This suggests that our three-body operators
transform as flavor octets in the limit $r=1$. 

Finally, if three-body operators are considered
and SU(3) symmetry is explicitly broken, 
only the (64,3), (64,2), (35,2), and (27,2) relations 
with isospin $I \ge 2$ are satisfied.
These can be obtained from linear combinations of our 
Eqs.(\ref{rel6a}-\ref{rel6f}). The restriction 
to quadrupole operators that are linear in the quark charge
implies that one has  only $I=0$ and $I=1$ 
operators, which cannot affect $I \ge 2$ combinations.
Lebed's relations involve more
quadrupole moments than our relations
because isospin symmetry is not assumed in Ref.~\cite{Leb95}.

\subsection{Numerical results} 
After neglecting three-body operators $(C=0)$, one can express the 18 
quadrupole moments in terms of only one constant, $B$, which we have 
determined from the empirical neutron charge radius 
$r_n^2=-0.113(3)$ fm$^2$~\cite{Kop95}. 
Numerical values 
are listed in Tables~\ref{quadmonum} and \ref{transquadnum} for the cases
without ($r=1$) and with ($r=0.6$) flavor symmetry breaking. 
The electric quadrupole moments of the charged baryons are of the same 
order of magnitude as $r_n^2$, while those of the 
neutral baryons are considerably smaller.
We expect that the inclusion of three-quark operators will not  
change the sign and the order of magnitude of the numerical results
obtained. 

With the cubic type of SU(3) symmetry breaking 
we obtain similar numerical values.
For the quadrupole moments of most charged baryons and all
transition quadrupole moments differences between both types of 
symmetry breaking are of the order of 20$\%$. 
Larger differences occur for $Q_{\Sigma^{* 0}}$, $Q_{\Xi^{* 0}}$, 
and $Q_{\Omega^-}$. Here, 
the details of how flavor symmetry is broken are important. 
For example, from Eq.(\ref{rel8e}) we obtain $Q_{\Omega^-}=0.024$ fm$^2$ 
compared to $Q_{\Omega^-}=0.041$ fm$^2$ following from Eq.(\ref{rel7d}).
Nevertheless, if a quadrupole moment of this order of magnitude
is measured, one could distinguish between various theoretical 
approaches~\cite{Oh94}.
\begin{table}[htb]
\begin{center}
\begin{tabular}{ l   r  r  r  }
\hline
\hline
& ${Q}(r=1)$ & $Q$(quadratic)  & $Q$(cubic) \\[0.15cm] \hline
$\Delta^{-}$	  &  $ 0.113$  & $ 0.113$  & $ 0.113 $    \\
$\Delta^{0}$	  &  0         &     0     &  0        \\
$\Delta^{+}$	  &  $-0.113$  & $-0.113$  &  $-0.113$    \\
$\Delta^{++}$	  &  $-0.226$  & $-0.226$  & $-0.226$   \\
$\Sigma^{\ast -}$ &  $ 0.113$  & $ 0.083$  & $0.074 $   \\
$\Sigma^{\ast 0}$ &    0       & $-0.008$  & $-0.017 $  \\
$\Sigma^{\ast +}$ &  $-0.113$  & $-0.105$  & $-0.107$    \\ 
$\Xi^{\ast -}$    &  $ 0.113$  & $ 0.059$  &  $0.044 $  \\
$\Xi^{\ast 0}$    &    0       & $-0.009$  & $-0.023$  \\
$\Omega^-$	      &  $ 0.113$  & $ 0.041$  & $0.024 $\\   
\hline
\hline
\end{tabular} 
\caption[C2 moments]{
Numerical values for the quadrupole moments 
of decuplet baryons in units of [fm$^2$] 
according to the analytic expressions in 
Table~\ref{quadmo} with $B=r_n^2/4$ and $C=0$. 
The experimental neutron charge radius~\cite {Kop95},  
$r_n^2=-0.113(3)$ fm$^2$, and the SU(3) symmetry breaking 
parameter~\cite{Buc00}, $r=0.6$, are used as input values. }
\label{quadmonum}
\end{center}
\end{table}
\begin{table}[htb]
\begin{center}
\begin{tabular}{ l  r r r  }
\hline
\hline
  & ${Q}(r=1)$ & $Q$(quadratic) & $Q$(cubic)  \\[0.15cm] \hline
$p\rightarrow \Delta^+$ & $-0.080$     & $-0.080$   & $-0.080$    \\
$n\rightarrow \Delta^0$ &  $-0.080$    & $-0.080$   & $-0.080$  \\
$\Sigma^- \rightarrow \Sigma^{\ast -}$ & $0$        & $0.021$ & $0.028$   \\
$\Sigma^0 \rightarrow \Sigma^{\ast 0}$ & $-0.040$   & $-0.035$ & $-0.028$  \\
$\Lambda^0 \rightarrow \Sigma^{\ast 0}$& $-0.069$   & $-0.042$ & $-0.042$  \\
$\Sigma^+ \rightarrow \Sigma^{\ast +}$ & $-0.080$   & $-0.090$ & $-0.084$\\
$\Xi^- \rightarrow \Xi^{\ast -}$       & 0 & $0.013$ & $0.014$   \\
$\Xi^0 \rightarrow \Xi^{\ast 0}$       & $-0.080$  & $-0.035$ & $-0.034$ \\
\hline
\hline
\end{tabular} 
\caption[C2 Transition moments]{Numerical values for the 
octet-decuplet transition quadrupole moments  in units of [fm$^2$] according
to the analytic expressions in Table~\ref{transquad} with 
$B=r_n^2/4$ and $C=0$.
The experimental neutron charge radius~\cite {Kop95},  
$r_n^2=-0.113(3)$ fm$^2$, and the SU(3) symmetry breaking 
parameter~\cite{Buc00}, $r=0.6$,
are used as input values. 
The experimental $N \to \Delta$ transition quadrupole moments extracted from 
the measured $E2/M1$  ratios  
  $Q^{exp}_{p\to \Delta^+}=-0.105(16)$ fm$^2$~\cite{Bla97}, and  
  $Q^{exp}_{p\to \Delta^+}=-0.085(13)$ fm$^2$~\cite{Bec97}
agree well with our results using $C=0$.} 
\label{transquadnum}
\end{center}
\end{table}

In the SU(3) limit, the quadrupole moments of the neutral
baryons are exactly zero. In addition, the transition moments involving the
negatively charged baryons are zero, because of $U$-spin conservation,
which forbids such transitions if flavor symmetry is exact~\cite{Lip73}. 
Furthermore, the sum of all decuplet quadrupole moments is zero in this limit.

\section{Summary} 
Morpurgo's general QCD parameterization method was used to relate the 
spectroscopic quadrupole moments of all members of the baryon 
decuplet and the corresponding octet-decuplet transition quadrupole moments. 
Our analysis includes two- and three-quark currents. Because the method 
relies mainly on the symmetries of QCD, our predictions are
to a large extent model-independent. 
The model-dependence resides in our approximate treatment of flavor-symmetry 
breaking. A more rigorous approach will mainly affect the numerically small
quadrupole moments.

We have found eleven relations between the  
quadrupole moments that are
independent of the way SU(3)-flavor symmetry is broken.  Among them
are four relations, which follow already from the assumption 
of SU(2) isospin symmetry of the strong interaction and the 
linearity of the electric quadrupole moment operator in the quark charge.
These should be very well satisfied in nature.
 
We have compared our results with a purely group theoretical analysis
of quadrupole moments by Lebed. Our theory reproduces all Lebed 
relations. In addition, we find that some of his relations are more general, 
and hold even with certain types of three-quark operators included.

Measurements of baryon quadrupole moments are difficult
but within reach. They are important for determining the geometric shape 
of baryons. For example, the $\Omega^-$ baryon 
lives sufficiently long to observe the  X-rays 
emitted when it is captured into an outer Bohr orbit of a heavy 
nucleus and cascades down to a state with a lower principal quantum
number~\cite{Ste73}. The X-ray frequency  depends on whether the 
charge distribution of the $\Omega^-$ is spherically symmetric or deformed.
If it is deformed, the additional interaction energy 
between its quadrupole moment and the electric field gradient of 
the nucleus will lead to a frequency shift proportional to $Q_{\Omega^-}$.  
It has also been suggested to measure the degree of longitudinal 
spin polarization of an 
initially transversely polarized $\Omega^-$ beam  when passing through a 
crystal~\cite{Bar93}.
The interaction between the crystal electric field and 
the $\Omega^-$ quadrupole moment results in a longitudinal polarization, 
which is again proportional to $Q_{\Omega^-}$. 

In the Primakoff reaction $ Y + Pb \to Y^* +Pb $, a high energy
octet hyperon $Y$ is inelastically scattered in the electromagnetic 
field of, e.g., a $Pb$  nucleus, and a decuplet hyperon $Y^*$ is produced 
in the final state~\cite{Sel98}. An octet-decuplet transition quadrupole 
moment will affect the $Y^*$ production rate and its size may be extracted
from the Primakoff cross section.  Transition quadrupole
moments can also be obtained from the cross section for kaon photoproduction.
In the cross section for 
$\gamma p \to K^+ \Sigma^{* 0} \to K^+ \Lambda^0 \gamma$
~\cite{Sch95} the radiative decay width for 
$\Sigma^{* 0} \to \Lambda^0 \gamma$ with its 
magnetic dipole (M1) and electric quadrupole (E2) contributions enters.
The E2 contribution is a measure of the transition quadrupole moment.

With the help of the present theory the experimentally inaccessible 
quadrupole moments can be obtained from those that can be measured, 
and the geometric shape of baryons 
can be calculated in various models~\cite{Hen01}.

\vspace{0.5 cm}

{\bf Acknowledegement}: E. M. H. thanks Professor Amand Faessler and the 
members of the Institute for Theoretical Physics at the University
of T\"ubingen for hospitality during his stay in July 1999,
when the idea for this work was conceived. He also thanks the Alexander
von Humboldt Foundation for a grant. A. J. B. thanks 
the Institute for Nuclear Theory of the University of Washington 
for hospitality during a visit in September 1999 
and some financial support. We thank G. Morpurgo and 
R. Lebed for helpful correspondence.

\vfill
\eject


\begin{thebibliography}{99}
\bibitem{comment1} 
Spin-flavor symmetry breaking manifests itself 
in a lifting of the baryon octet and decuplet mass degeneracy,
in a nonzero neutron charge radius, and in a deviation of baryon 
charge distributions from spherical symmetry as reflected by 
finite baryon quadrupole moments.
One can see this, e.g, from the quark model relation 
$r_n^2=-b^2 \, (M_{\Delta}-M_N)/M_N$ between the $\Delta$-$N$ 
mass splitting $M_{\Delta} - M_N$,  the radius of the spatial quark (mass)
distribution $b$, and the neutron charge radius $r_n^2$~\cite{Buc97}.
\bibitem{comment2} The additional factor $1/m_j$ in the cubic type symmetry
breaking is due to the intermediate quark propagator between the photon-quark
and gluon-quark vertices.
\bibitem{comment3} The same is true when ``cubic'' and 
``quadratic'' SU(3) symmetry breaking are used respectively
for two- and three-body operators.
\bibitem{comment4} The second line in the (27,0) 
relation of Ref.~\cite{Leb95} should have an overall minus sign.
\end{thebibliography}
\end{document}